\documentclass[submission,copyright,creativecommons]{eptcs}
\providecommand{\event}{WWV 2013\\Automated Specification and Verification of Web Systems\\9th International Workshop}
\providecommand{\titlerunning}{Proving Properties of Rich Internet Applications}
\providecommand{\authorrunning}{James Smith}
\usepackage{breakurl}             

\usepackage{common}

\newcommand*\circled[1]{\tiny\tikz[baseline=-3pt]{\node[shape=circle,draw,inner sep=1.25pt] (char) {#1};}\normalsize}

\Pifont{pzd}
\newcommand{\cmark}{\text{\ding{51}}}%

\makeatletter

\newcommand{\Rom}[1]{\expandafter\@slowromancap\romannumeral #1@}
\makeatother

\newcommand{\doc}{d\hspace{-0.075em}oc}		
\newcommand{\temp}{t\hspace{-0.075em}emp}
\newcommand{\diffs}{d\hspace{-0.075em}i\hspace{-0.125em}f\hspace{-0.25em}f\hspace{-0.125em}s}
\newcommand{\diffss}{d\hspace{-0.075em}i\hspace{-0.125em}f\hspace{-0.25em}f\hspace{-0.125em}ss}
\newcommand{\luid}{l\hspace{-0.075em}uid}
\newcommand{\empt}{/\hspace{-0.5em}\raisebox{-0.075em}{0}}
\newcommand{\empti}{/\hspace{-0.6em}\raisebox{-0.075em}{0}}

\title{Proving Properties of Rich Internet Applications}
\author{James Smith
\institute{Imperial College\\ London, United Kingdom}
\email{jecs@imperial.ac.uk}
}
\def\titlerunning{Proving Properties of Rich Internet Applications}
\def\authorrunning{James Smith}

\begin{document}

\maketitle

\setlength\belowcaptionskip{0pt}

\numberwithin{figure}{section}

\begin{abstract}
\noindent We introduce application layer specifications, which allow us to reason about the state and transactions of rich Internet applications. We define variants of the state/event based logic UCTL* along with two example applications to demonstrate this approach, and then look at a distributed, rich Internet application, proving properties about the information it stores and disseminates. Our approach enables us to justify proofs about abstract properties that are preserved in the face of concurrent, networked inputs by proofs about concrete properties in an Internet setting. We conclude that our approach makes it possible to reason about the programs and protocols that comprise the Internet's application layer with reliability and generality.
\end{abstract}

\section{Introduction}

A glance at the existing literature shows that the Internet's transport layer is well understood from a mathematical point of view, with the inductive analysis of the TLS protocol~\cite{DBLP:journals/tissec/Paulson99} being perhaps the best example. On the other hand, the application layer that sits above it encompases a plethora of languages and frameworks, and understanding seems to be incomplete. At the highest level, the behaviour of Internet programs and protocols has been analysed~\cite{DBLP:conf/csfw/AkhaweBLMS10,DBLP:conf/oopsla/MartinLL05,DBLP:journals/jsc/HemelGKV11}, often for security reasons, but nevertheless there appears to be much work still to do.

These days, Internet applications are made up of both server side and client side parts. These so called ``rich'' Internet applications are now so commonplace as to be the norm. Any formalism that helps us to understand them mathematically seems a worthy goal, therefore. A complete formal treatment would include a formalisation of JavaScript~\cite{DBLP:conf/popl/GardnerMS12} and much more, however, so here we attempt a loss lofty goal, giving formal specifications and proofs for that part of their behaviour related to communication over HTTP. To give an example, consider a ``shoot-em-up'' style game implemented in JavaScript and run in a web page. The functionality related to the unfolding of the gameplay would encompass game logic and the rendering of each scene, functionality not directly related to the Internet. Suppose the game includes a high score table, however, with high scores stored on the server. Maintaining this table would encompass issues relating to concurrency and communication over HTTP. It is the formal specifications and proofs for this kind of behaviour that is the subject of this paper.

Our motivation in studying this kind of behaviour is a broad understanding of the programs and protocols that comprise the Internet's application layer. With this in mind, this paper demonstrates that it is possible to formalise certain properties of rich Internet applications using a combination of mathematical models, transition systems, logics and related techniques. The approach is based on the conventional HTTP request-response mechanism whose participants are usually the web server on the one hand and the web browser on the other. Crucially, however, since communication between the client side part of an Internet application with its server side part, usually carried out by way of the XMLHTTPRequest object, uses the same HTTP request-response mechanism, our approach encompasses the behaviour of rich Internet applications, too.

\noindent Our contributions are therefore threefold. Firstly, we formalise HTTP transactions and Internet application state, and introduce the concept of transition rules which describe the relation between the two. This constitutes our model of the Internet's application layer. Secondly, we define variants of the logic UCTL*~\cite{uctlstar}, whose semantics are defined over the variants of Kripke transition systems~\cite{DBLP:conf/sas/Muller-OlmSS99} that these transition rules generate. We use these logics to express properties of these transition systems. Thirdly, in the case study we present a formal treatment of a distributed, rich Internet application and prove properties about the information it stores and disseminates. Our approach is a general one in the sense that we do not rely on a particular framework or programming paradigm. The properties we prove are ones that could be exhibited by any HTTP-based Internet program or protocol.

\section{HTTP transactions and Internet application state}

To the uninitiated an Internet application may appear to persist. A web page loads and then remains in the browser in much that same way that a desktop application remains in front of the user. The real situation is somewhat different, however. The presence of the web page may give the impression of permanence but in reality the instance of the server part of the application that produced it is likely to be destroyed the moment the web page is produced. By contrast, the client side part of a rich Internet application does persist, it can communicate with the server side part alongside the browser and it has access to and can amend shared information stored by the browser. To complicate the picture still further, the freedom that the browser admits means that user interactions cannot be guaranteed to occur sequentially in relation to the unfolding of the application state. The user may open new tabs, refresh the page, type URLs directly into the address bar or, most notoriously, use the forward and back buttons. The effects of these actions are complex to discern to say the least. 

Rather than attempting to model this cat's cradle, we take a simplified approach. It begins with the working of the HTTP protocol and in particular the request-response pair, which we call an HTTP transaction. Although it is possible for the request-response pair to be broken, the connection may fail or the server may fault, for example, in our approach it will be considered inviolable. Many transactions may happen, building up the complete picture, so we therefore take account of all possible transactions. Although if we make the assumption that no two transactions can occur at exactly the same time then they can be given a temporal ordering, such an ordering is of little practical use since it is extremely difficult if not impossible to establish any causal link between successive transactions. It is best therefore to think of the transactions to begin with as floating around in a kind of ``soup'', with no connections between them at all.

At this point we make two observations. The first is that HTTP transactions can happen as a result of requests from both the browser and the client side part of the application, and therefore this soup may contain both types. The second is that no account is taken of the cause of each transaction. Each may happen as a result of the user clicking a link; pressing the back button; manually typing an address into the address bar; the client side part of the application making a call via the XMLHTTPRequest object; the list is endless. By eschewing these causes we avoid the trap of failing to properly account for them. Furthermore, we do not have a model of browser architecture or attempt to discern the subtle interplay of user interactions with iframes, different tabs and so on. And if we lose something in taking this line what we gain is a complete model of what remains, namely the behaviour of the application as evinced by its transactions over HTTP.

This soup of transactions on its own is of little use. To begin with it takes no account of the state of the application, however this is defined. And some causal links must be established, some additional structure needs to be added. To do this we include the states of the application in our approach, and associate them with transactions. More precisely, we say that if an application is in a certain state before an HTTP request, its state may change after the HTTP response. We call this process a transition, with that part of the transition visible over HTTP being precisely the transaction.

Our approach therefore consists of modelling all possible application states together with all possible transactions between them. We do not consider the causes of requests, and we assume that any request can happen at any time. This approach has precedents in the literature on the verification of network protocols, for example~\cite{DBLP:conf/isola/RavnSV10}. In such cases there is no user interface to be modelled, and verification consists of exhaustively ``covering all the bases'' with regard to all possible transactions and their effects on application state. And this approach also characterises some of the most stringent web security requirements in industry, including~\cite{PCI-DSS}, where no assumptions are made about the types of requests that can be made in an ``anything can happen'' analysis. Our approach is a natural extension of these approaches.

\section{A formal model and examples}

In this section we define a formal model of HTTP transactions and Internet application state as well transition rules, which describe the relation between the two. We then define two example applications based on this model. The first application uses cookies to enforce user flow through a site and demonstrates the persistence of application state in the browser. The second application maintains a server state in order to track visitors to a site. 

The HTTP protocol is a request-response protocol that admits communication between a client, typically a web browser or a JavaScript program running in a browser, and a server, typically an instance of a  script or program running on a web server. Both requests and responses consist of a number of headers together with a body. The request often contains a reference, called a uniform resource locator or URL for short, to a resource to be retrieved by the server. Scripts and programs on the server can also be invoked in this manner. The headers communicate cookies in both directions, alongside other information. 

Cookies consist of name-value pairs of textual information and allow stateful communication between clients and servers. They are initially passed by the server to the client in response to a request. They are then stored in the browser and passed back to the server in subsequent requests. Usually the instance of a script or program that produces any response is destroyed the moment that reponse is produced and therefore cookies allow subsequent instances to recover information specific to a particular client. Typically an identifier unique to the client is stored in a cookie, which allows information specific to that client to be retrieved from a database on the server. This leads to the common misconception that the identifier itself persists there. Aside from this, state not related to any particular client may be held on the server. Also a JavaScript program running in the browser may hold state apart from the browser's cookie store.

We model cookies as atomic entities rather than name-value pairs, taking them from a set of cookies $\mathscr{C}$, with $c$, $c'$ ranging over $\mathcal{C}=\mathscr{P}(\mathscr{C})$. Since the browser can effectively be instructed to both add and remove cookies from its store, we model them as being signed when passed to the browser, so we have a set of signed cookies $\mathscr{C}^\pm=\{+x|x\in\mathscr{C}\}\cup\{-x|x\in\mathscr{C}\}$, with $c^\pm$ ranging over $\mathcal{C}^\pm=\mathscr{P}(\mathscr{C}^\pm)$. Global states are tuples containing browser's cookie store, which from now on we refer to as the browser's state, together with the client side and server side states:
\[
(c,j,s)\in\mathcal{C}\times\mathcal{J}\times\mathcal{S}
\]
Requests consist of a set of cookies $c\in\mathcal{C}$ together with a URL $u$ taken from a set of URLs $\mathcal{U}$. Responses consist of a set of signed cookies $c^\pm\in\mathcal{C}^\pm$ together with a body $b$ taken from a set of bodies $B$. We may refine requests and responses further as required, adding additional request headers, for example, or parameterising the response bodies. A transaction consists of a request and response in one tuple, which labels a transition between global states:

\begin{figure}[H]
\centering
\includegraphics[scale=1]{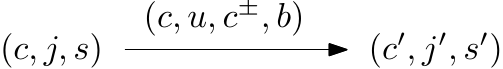}
\end{figure}

\noindent Note that the browser's state $c$ is echoed directly in the request, and therefore there is some redundancy in this formalism. We keep this, however, so that the transaction represents that part of the transition visible over HTTP in its entirety. To continue, when the browser receives signed cookies, it amends its state according to the following generic rule:
\[
c'=c\cup\{x\in\mathscr{C}\;|\;+x\in c^\pm\}\backslash\{x\in\mathscr{C}\;|\;-x\in c^\pm\}
\]
\noindent Transitions are governed by transition rules, which, given an initial state and request, determine both the response and final state:
\[
(c,j,s)\xrightarrow[\;\;c^\pm,b\;\;]{c,u}(c',j',s')
\]
Transition rules are typically obtained from static analysis. This process could be automated or might be an informal but nonetheless informed approximation, as in case study given later. In the case of the examples that follow, however, the transition rules are formally stated and effectively define the applications in question. 

\subsection{The `agreement' example}

When visiting a website, a user must first agree to the use of cookies, then agree to the terms and conditions, and only then are they given access to the welcome page. No assumptions can be made about the order in which the pages are requested, however. In order to enforce the correct flow, two nonce cookies are set when the user agrees to the relevant missives. An error page is shown if the user requests the pages out of order, say, or bookmarks a page in order to come back to it at a later stage when the cookies have been deleted. 

The user agrees both to the use of cookies and to the terms and conditions by way of forms presented on the first two pages. In both cases, only if the user ticks the checkbox and then submits the form are they allowed to continue. Upon successful submission of the forms nonce cookies are set, and it is the presence or otherwise of these cookies which determines which pages can be shown. No distinction is made between a page requested via the address bar and form submissions with the checkboxes left unticked, in which case the input is treated as invalid and the relevant page is shown again. 

\begin{figure}[h]
\centering
\includegraphics[scale=1]{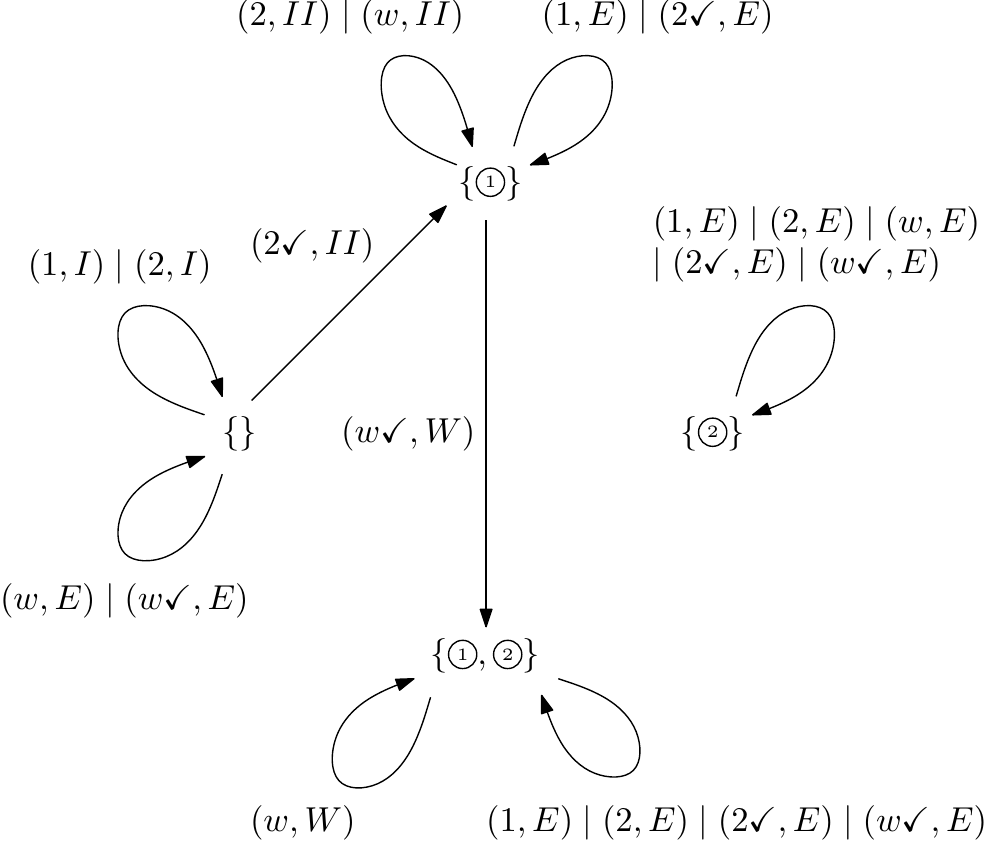}
\caption{The `agreement' application transition system}
\label{ts_agreement}
\end{figure}

We model the two nonce cookies, plus the URLs that are used to request the three pages. To model valid form submissions, two additional URLs with ticks are added. We model the four page bodies returned, the first two containing the forms, then the welcome and error pages:
\[
\mathcal{C}=\mathscr{P}(\{\circled{1},\circled{2}\})\quad\mathcal{U}=\{1,2,w,2\cmark,w\cmark\}\quad\mathcal{B}=\{\Rom{1},\Rom{2},W,E\}
\]

\noindent The application's behaviour is defined by the following set of transition rules:
\[
\begin{array}{ccccc}
\{\}\xrightarrow[\{\}\;\Rom{1}]{\{\}\;1\;|\;2}\{\}&{\hspace{1em}}&
\{\circled{1}\}\xrightarrow[\{\}\;\Rom{2}]{\{\circled{1}\}\;2\;|\;w}\{\circled{1}\}&{\hspace{1em}}&
\{\circled{1},\circled{2}\}\xrightarrow[\{\}\;W]{\{\circled{1},\circled{2}\}\;w}\{\circled{1},\circled{2}\}
\end{array}
\]
\[
\begin{array}{ccc}
\{\}\xrightarrow[\{+\circled{1}\}\;\Rom{2}]{\{\}\;2\cmark}\{\circled{1}\}&{\hspace{1em}}&
\{\circled{1}\}\xrightarrow[\{+\circled{2}\}\;W]{\{\circled{1}\}\;w\cmark}\{\circled{1},\circled{2}\}
\end{array}
\]
\[
\begin{array}{ccccc}
\{...\}\xrightarrow[\{\}\;E]{\{\circled{1}\}\;|\;\{\circled{2}\}\;|\;\{\circled{1},\circled{2}\}\;1}\{...\}&{\hspace{1em}}&
\{...\}\xrightarrow[\{\}\;E]{\{\circled{2}\}\;|\;\{\circled{1},\circled{2}\}\;2}\{...\}&{\hspace{1em}}&
\{...\}\xrightarrow[\{\}\;E]{\{\}\;|\;\{\circled{2}\}\;w}\{...\}
\end{array}
\]
\[
\begin{array}{ccc}
\{...\}\xrightarrow[\{\}\;E]{\{\circled{1}\}\;|\;\{\circled{2}\}\;|\;\{\circled{1},\circled{2}\}\;2\cmark}\{...\}&{\hspace{1em}}&
\{...\}\xrightarrow[\{\}\;E]{\{\}\;|\;\{\circled{2}\}\;|\;\{\circled{1},\circled{2}\}\;w\cmark}\{...\}
\end{array}
\]
The first row represents the occasions when the user requests the correct page or submits a form in the correct order but without ticking the checkbox. The second row represents the occasions when the user correctly submits a form and a nonce cookie is set. The third and fourth rows represent the remaining occasions, on all of which error pages are shown. Note that both the client side and server side states are missing from these rules, since they are not present in this example. Note also that for brevity's sake, the browser states are sometimes replaced by ellipses with the understanding that they can be inferred from the headers. Recall that the browser's state is echoed directly in the request. Finally, note that vertical bars allow several rules to be combined into one. Figure~\ref{ts_agreement} shows the resulting transition system, which in this example is no more than a diagrammatic representation of the transition rules. The cookie headers are omitted, since they can be inferred from the initial and final states. Note that we also use the vertical bar rather than set notation, in keeping with the transition rules.

\subsection{The `visitors' example}

A website maintains a counter on a landing page which is incremented every time a user requests that page for the first time. No cookies can be used and so in order to determine whether a request has originated from a new user, the request IP address is checked against the server state. If the request IP address cannot be found, the counter is incremented and the new IP address is added to the server state. 

\begin{figure}[H]
\centering
\includegraphics[scale=1]{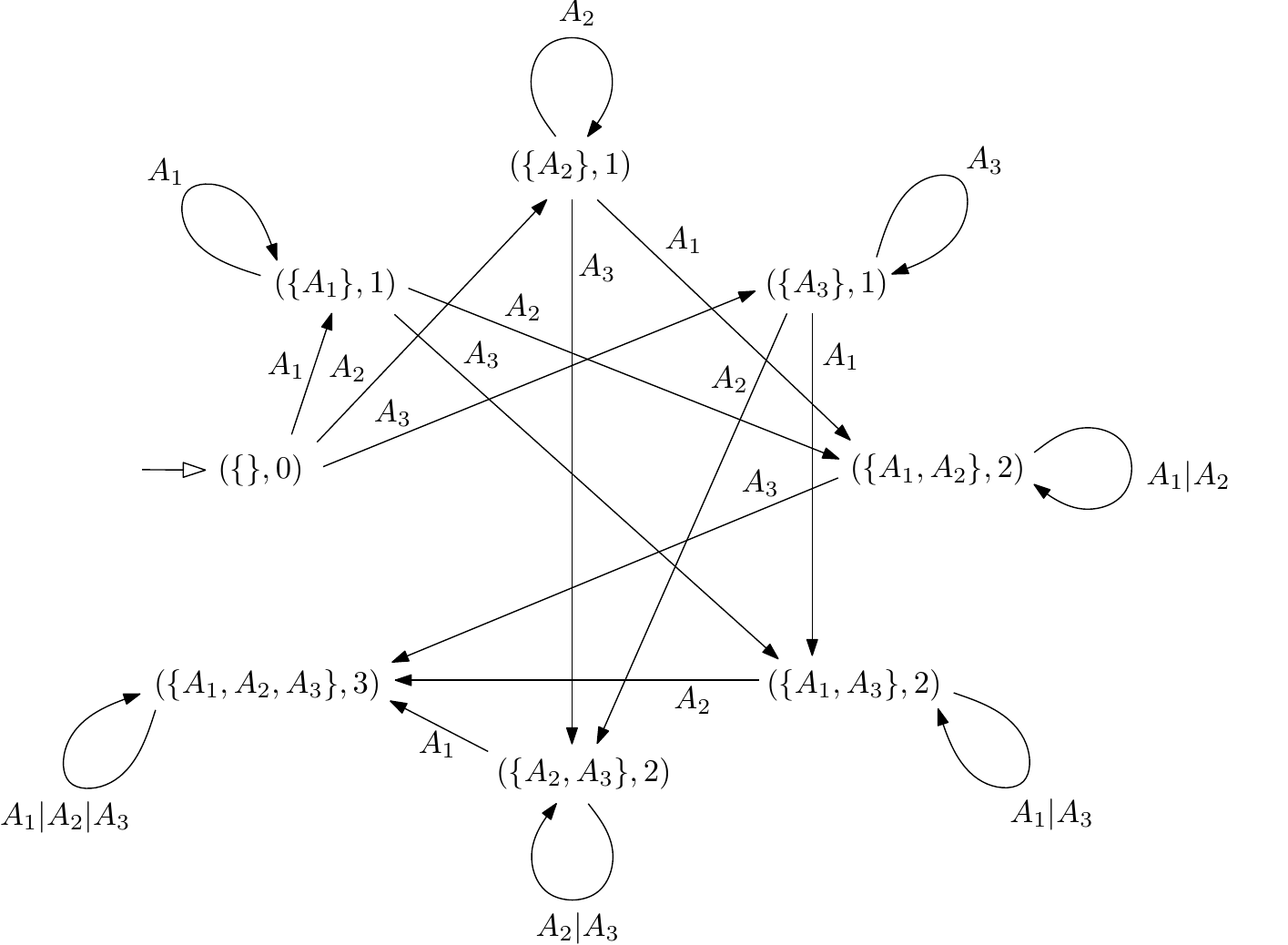}
\caption{The `visitors' application transition system}
\label{ts_visitors}
\end{figure}

\noindent We model the lack of cookies, a single URL, a set of IP addresses and a countable set of page bodies. The server state is modelled as a subset of the set of the IP addresses paired with the states of the counter, which are precisely the natural numbers $\mathbb{N}$. In this case the application also has an initial state:
\[
\mathcal{C}=\mathscr{P}(\{\})\quad\mathcal{U}=\{U\}\quad\mathcal{A}=\{A_1,A_2,\cdots,A_N\}\quad\mathcal{B}=\{B(n)\;|\;n\in\mathbb{N}\}
\]
\[
(a,n)\in\mathcal{S}\subseteq \mathscr{P}(\mathcal{A})\times\mathbb{N}\quad(\{\},0)\in\mathcal{S}
\]
The application's behaviour is defined by the following two transition rules for $1\leqslant i\leqslant N$:
\[
\begin{array}{ccc}
(a,n)\xrightarrow[B(n)]{\;\;U\!\!,\,\,A_i\;\;}(a,n)\quad A_i\in a
&
\quad\quad\quad\quad\quad
&
(a,n)\xrightarrow[B(n+1)]{U\!\!,\,\,A_i}(a\cup\{A_i\},n+1)\quad A_i\notin a
\end{array}
\]
Whereas the IP address is communicated alongside the URL in the request, on the other hand the counter is communicated as part of the page body. The returned page is essentially parameterised by the counter, therefore, reflecting the fact that the counter is embedded in the body of the page by some means. Note that the browser and client side states are missing from these rules, since neither is not present in this example. Note also that since there is no browser state, no cookies are communicated in the request and response parts of the rules. Strictly speaking these two transition rules are not single rules but rule schemas defining two countable sets of rules for each $(a,n)\in \mathscr{P}(\mathcal{A})\times\mathbb{N}$. Figure~\ref{ts_visitors} shows the resulting transition system in the case when $N=3$. The initial state is highlighted on the left with a white filled arrow. Since there is only one request URL and one returned page, these are omitted. It is also assumed that the correct value of the counter is echoed in the returned page and therefore this information is not included. Note again that we use the vertical bar rather than set notation for the transition labels.

\section{A treatment of the examples using temporal logics}

The two transition systems defined in the previous section are, with small differences, instances of Kripke transition systems, with information on both the states and transitions. In order to express properties of the applications in question we therefore use a variant of UCTL*~\cite{uctlstar}, a state/event based logic whose semantics are defined over these types of transition systems. In what follows we give the standard definition of Kripke transition systems~\cite{DBLP:conf/sas/Muller-OlmSS99}, together with the syntax and semantics of UCTL*, and then define variants of both for our purposes.

\begin{definition}A Kripke Transition System is a tuple $(S,Act,{\longrightarrow},AP,\mathcal{L})$ where:
\begin{itemize}
\item $S$ is a set of states ranged over by $s$,$s_0$ and $s_1$,
\item $Act$ is a set of actions, with $2^{Act}$ ranged over by $\alpha$,
\item ${\longrightarrow}\subseteq S\times 2^{Act}\times S$ is the transition relation with $(s_0,\alpha,s_1)\in{\longrightarrow}$,
\item $AP$ is a set of atomic propositions ranged over by $p$,
\item $\mathcal{L}:S\times AP{\longrightarrow}\{true,false\}$ is an interpretation function that associates a value of $true$ or $false$ with each $p\in AP$ for each $s\in S$,
\item For any two transitions, $(s_0,\alpha_0,s_1),(s_0,\alpha_1,s_1)\in{\longrightarrow}\Rightarrow\alpha_0=\alpha_1$.
\end{itemize}
\end{definition}

\noindent Note that since transitions carry sets of actions and not just one, there is no silent action, with the empty set $\{\}$ being considered silent instead. Note also that we limit the number of transitions between any two states in any one direction to at most one.

Paths are sequences of transitions where the final state of one transition equals the initial state of the next, if there is one. They are ranged over by $\sigma,\sigma'$ and $\sigma''$. Maximal paths are either infinite or their last state has no outgoing transitions. For the set of maximal paths starting at state $s$ we write $\mu path(s)$. For the initial and final states of the first transition of a path $\sigma$ we write $\,_\mathsf{S}\sigma$ and $\sigma_\mathsf{S}$, respectively, and we write $\sigma_\mathsf{T}$ for the set of actions of the first transition of a path $\sigma$. A suffix $\sigma'$ of a path $\sigma$ is a path such that $\sigma=\sigma''\sigma'$ for some possibly zero length path $\sigma''$. A proper suffix $\sigma'$ of a path $\sigma$ is a path such that $\sigma=\sigma''\sigma'$ for some non-zero length path $\sigma''$. We write $\sigma\leqslant\sigma'$ when $\sigma'$ is a suffix of $\sigma$ and $\sigma<\sigma'$ when $\sigma'$ is a proper suffix of $\sigma$.

\begin{definition}The syntax UCTL* is:
\[
\phi::=p\;|\;\neg\phi\;|\;\phi\wedge\phi'\;|\;\exists\pi\quad\pi::=\phi\;|\;\neg\pi\;|\;\pi\wedge\pi'\;|\;X\pi\;|\;X_a\pi\;|\;\pi\; U\pi'
\]
Here $\phi$ is a state formula and $\pi$ a path formula.
\end{definition}

\begin{definition} 
The semantics of UCTL* is:

\begin{itemize}
\item $s\models p$ iff $\mathcal{L}(s,p)=true$
\item $s\models\neg\phi$ iff $s\nmodels\phi$
\item $s\models\phi\wedge\phi'$ iff $s\models\phi$ and $s\models\phi'$
\item $s\models\exists\pi$ iff $\exists\sigma\in\mu path(s):\sigma\models\pi$
\item $\sigma\models\phi$ iff $\,_\mathsf{S}\sigma\models\phi$
\item $\sigma\models\neg\pi$ iff $\sigma\nmodels\pi$
\item $\sigma\models\pi\wedge\pi'$ iff $\sigma\models\pi$ and $\sigma'\models\pi'$
\item $\sigma\models X\pi$ iff $\exists(s,\alpha,s')\sigma'':\sigma=(s,\alpha,s')\sigma'',\sigma''\models\pi$
\item $\sigma\models X_a\pi$ iff $\exists(s,\alpha,s')\sigma'':a\in\alpha,\sigma=(s,\alpha,s')\sigma'',\sigma''\models\pi$
\item $\sigma\models\pi\;U\pi'$ iff $\exists\sigma'\geqslant\sigma:\sigma'\models\pi',\forall\sigma'':\sigma\leqslant\sigma''<\sigma':\sigma''\models\pi$
\end{itemize}

\end{definition}

\noindent We now define a variant of UCTL* for our purposes, replacing atomic propositions with propositions about typed variables.

\begin{definition}
Let $x$ be a typed variable. The type of $x$, written $\tau_x$, is the set of its permitted values. A set of typed variables is written $X$.
\end{definition}

\begin{definition}
An interpretation of a set of typed variables $X$ is a function $I$ from $X$ to $\bigcup\{\tau_x|x\in X\}$ such that $I(x)\in\tau_x$ for each $x\in X$. The set of all interpretations for a set of typed variables $X$ is written $\mathbb{I}$.
\end{definition}

\begin{definition}An (amended) Kripke Transition System a tuple $(S,Act,\longrightarrow,X,\mathcal{L})$ where the set of atomic propositions $AP$ and interpretation function $\mathcal{L}$ replaced with the following:
\begin{itemize}
\item $X$ is a set of typed variables,
\item $\mathcal{L}:S{\longrightarrow}\mathbb{I}$ is an interpretation function that maps every state to an interpretation.
\end{itemize}
\end{definition}

\noindent Action formulae remain the same, however state formulae must be amended in order to accomodate propositions of the form $x=x'$ where $x$ and $x'$ are of the same type, or $x=v$ where $v\in \tau_x$. In the spirit of~\cite{DBLP:conf/fsttcs/BohnDGHL98}, we also introduce a set $K$ of typed global variables, extending the syntax to include propositions of the form $x\leqslant k$ where $k\in K$, $x$ and $k$ are of the same type, the type is a numerical type that supports inequalities and the value of $k$ is in $\tau_x$. 

\begin{definition}
The (amended) syntax of UCTL* state formulae $\phi$ is:
\[
\phi\coloncolonequals true\;\;|\;\;x=v\;\;|\;\;x=x'\;\;|\;\;x\leqslant k\;\;|\;\;\neg\phi\;\;|\;\;\phi\wedge\phi'\;\;|\;\;\exists\pi
\]
\end{definition}

\begin{definition}
The (amended) semantics of UCTL* state formulae $\phi$ is:
\begin{itemize}
\item $s\models x=v$ iff $\mathcal{L}(s)(x)=v$
\item $s\models x=x'$ iff $\mathcal{L}(s)(x)=\mathcal{L}(s)(x')$
\item $s\models x\leqslant k$ iff $\mathcal{L}(s)(x)\leqslant k$
\end{itemize}
The relation $s\models p$ is dropped with the other relations remaining.
\end{definition}

\noindent If $x$ has a numerical type, we also permit expressions of the form $x=v+1$ with the obvious semantics. Unlike~\cite{DBLP:conf/fsttcs/BohnDGHL98}, which introduces typed constants, we might close formulae containing typed global variables under existential or universal quantification. If they are not closed in such a way, we consider them constant. We also quantify over sets of constants. 

Finally, we derive the ``eventually'' $F$ and ``always'' $G$ operators~\cite{Emerson:1983:SNR:567067.567081} in the usual way and add a $T$ operator, the latter being a convenient shorthand for combining path formulae satisfied immediately and ``nexttime'' on a path:

\[
F \pi'=true_{true}U_{true}\pi'\quad\quad G\pi=\neg F\neg\pi\quad\quad\pi T_\chi\pi'=\pi\wedge X_\chi\pi'
\]

\subsection{The `agreement' example}

Here $X=\{c\}$ and $\tau_c=\{\{\},\{\circled{1}\},\{\circled{2}\},\{\circled{1},\circled{2}\}\}$. For the sake of readability, formulae $x=v$ are shortened to just $v$. For example, $c=\{\circled{1}\}$ is written $\{\circled{1}\}$. The ideal user journey is the following path:

\begin{figure}[H]
\centering
\includegraphics[scale=1]{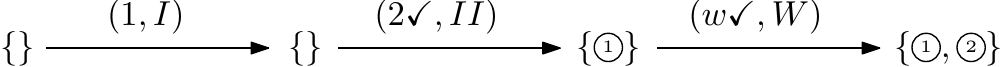}
\end{figure}

\noindent It is easy to produce a path with the same requests but the wrong responses, however:

\begin{figure}[H]
\centering
\includegraphics[scale=1]{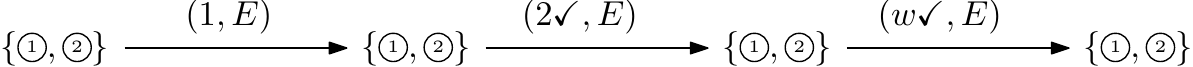}
\end{figure}

\begin{lemma}The following two formulae can be satisfied:

\[
\exists(\{\}T_{(1,\Rom{1})}\{\}T_{(2\cmark,\Rom{2})}\{\circled{1}\}T_{(w\cmark,W)}\{\circled{1},\circled{2}\})\quad\quad\exists(X_{(1,E)}X_{(2\cmark,E)}X_{(w\cmark,E)}true)
\]

\begin{proof} 
These follow immediately from inspecting figure~\ref{ts_agreement}, which includes the above paths.
\end{proof}
\end{lemma}

\subsection{The `visitors' example}

Here $X=\{a,n\}$, $\tau_a=\mathscr{P}(\mathcal{A})$, $\tau_n=\mathbb{N}$ and $K=\{N_1,N_2,N\}$ where $N_1,N_2,N\in\mathbb{N}$ and $N_1,N_2\leqslant N$. We continue to use an ordered pair $(a,n)$ for the state rather than a set.
\begin{lemma}
\label{lemma_state_initial}
The only state with $n=0$ and $a=\{\}$ is the initial state.
\begin{proof}
Only the second rule can be applied to the initial state, and there is no rule for subsequently decreasing $n$ or reducing $a$.
\end{proof}
\end{lemma}
\begin{lemma}
\label{lemma_state_form}
Apart from the initial state, all reachable states are of the form $(a\cup\{A_i\},n+1)$, where $A_i\notin a$, and are reached by way of a transition from state $(a,n)$.
\begin{proof}
The first transition rule leaves states unchanged, therefore if a state is not the initial state it must be reached by the application of the second transition rule to some state, say $(a,n)$, and therefore must be of the form $(a\cup\{A_i\},n+1)$.
\end{proof}
\end{lemma}
\begin{lemma}
For any reachable state $(a,n)$, $|a|=n$.
\begin{proof}
\label{lemma_a_equals_n}
By induction on $n$. For $n=0$ we know from lemma~\ref{lemma_state_initial} that the only state with $n=0$ is the initial state. Assume now that for all states $(a,k)$ we have $|a|=k$. Consider some other, reachable state with $n=k+1$ which by lemma~\ref{lemma_state_form} we can write as $(a\cup\{A_i\},k+1)$, reachable from the state $(a,k)$. By the induction hypothesis $|a|=k$ and therefore $|a\cup\{A_i\}|=k+1$.
\end{proof}
\end{lemma}
\begin{lemma}
For any reachable state $(a,n)$, $n\leqslant N$. Formally, $\forall G(n\leqslant N)$ is valid.
\begin{proof} 
We simply observe that $n=|a|\leqslant|\mathcal{A}|=N$.
\end{proof}
\end{lemma}
\noindent We omit the proof that there is a path that leads to a state $(a,n)$ for any $a\in \mathscr{P}(\mathcal{A})$ but note that this fact, in tandem with lemma~\ref{lemma_a_equals_n}, defines the set $\mathcal{S}$ of all reachable states.
\begin{lemma}
The value of the counter returned to the user is no greater than $N$.
\begin{proof}
For the first rule, the response is $B(n)$ with $n\leqslant N$. For the second rule, the response is $B(n+1)$ but $n=|a|$ by lemma~\ref{lemma_a_equals_n}, $a\subset\mathcal{A}$ and since $|\mathcal{A}|=N$, $n<N$ and therefore $n+1\leqslant N$.
\end{proof}
\end{lemma}
\begin{lemma}
No user can increment the counter twice. Formally, the following is not satisfiable:

\[
\exists A_i,N_1,N_2:\exists(((n=N_1)T_{A_i}(n=N_1+1))\wedge F((n=N_2)T_{A_i}(n=N_2+1)))
\]

\begin{proof}
Suppose that the formula is satisfied. The path in question must be of the following form:

\begin{figure}[H]
\centering
\includegraphics[scale=1]{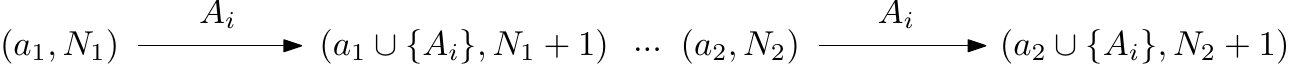}
\end{figure}

\noindent We first note that $a_1\cup\{A_i\}\subseteq a_2$, since all transitions are governed by two transition rules, one of which leaves the set $a$ of any state $(a,n)$ alone and the other of which adds an element. Then we note that in order for the second rule to be applied at the end of the path we must have $A_i\notin a_2$. However, $a_2\supseteq a_1\cup\{A_i\}$ and $A_i\in a_1\cup\{A_i\}$ implies $A_i\in a_2$, a contradiction.
\end{proof}
\end{lemma}

\vspace{-0.5em}

\section{The QuICDoc case study}

QuICDoc~\cite{quicdoc} is a small, distributed, rich Internet application that uses the Concur algorithm~\cite{concur} to merge concurrent changes to a shared document. A user simply types into a textarea and a client-side part of the application communicates the changes, what are called ``diffs'', to the server side part, which disseminates them to all the other client side parts after amending them in such as way as to ensure that they can be merged consistently. In~\cite{concur} it is proved that this is possible but no account is taken of the effects of implementing the algorithm in an Internet setting. We bridge that gap in this section.

To begin with we rule out concurrency effects at all levels bar the application level. QuICDoc is executed on Node.js~\cite{nodejs}, which uses an event-driven, non-blocking I/0 model, therefore we rule out concurrency effects at the system level without further ado. At the network level, once initialised the client side part of QuICDoc makes two repeated calls to the server side part. These are synchronous calls, opening a connection and providing a callback method. We mitigate against the chance of one callback interrupting another by introducing a shared \textsf{gUpdate} variable that stops one method making a call if the other method's callback is pending. The relevant parts of the client side code are shown immediately below:

\vspace{0.5em}

\begin{lstlisting}[basicstyle={\sffamily\footnotesize},frame=single,framesep=3pt,xleftmargin=3pt,xrightmargin=3pt,frameround=tttt]
var gUpdate;
var fGetDoc = function () {
    gUpdate = false; 
    fGetDiffs();
}
var fGetDiffs = function () {
    if ( gUpdate ) { setTimeout( "fGetDiffs()", 1000 ); return; }
    gUpdate = true;
}
var fPutDiffs = function () {
    if ( gUpdate ) { return; }
    gUpdate = true;
}
\end{lstlisting}

\vspace{0.5em}

\noindent Note that in order to avoid race conditions, the \textsf{fPutDiffs()} method exits if the \textsf{fGetDiffs()} callback is pending, meaning that the diffs generated by the user input are discarded. The next user input will generate fresh diffs with no information being lost, however. On the other hand the \textsf{fGetDiffs()} method is rescheduled if the \textsf{fPutDiffs()} callback is pending. Use of a shared variable to enforce asynchronicity in the face of timers, the event model and the like may seem na\"ive, however we maintain that it has the desired effect in practice. Moving on, at what we call the application level, concurrency effects are mitigated by the Concur algorithm.

\noindent In modelling the implementation of this algorithm we consider just two clients, defining the global state as a tuple $(j_1,j_2,s)\in\mathcal{J}_1\times\mathcal{J}_2\times\mathcal{S}$ encompassing both client side states alongside the server side state. In order to relate the formal model to the implementation we define states as named tuples with values associated with the relevant heap variables, thus $j_i(\mathsf{gUid},\mathsf{gWorkingDoc},\mathsf{gTempDoc})$  for $i=1,2$ together with $s(\mathsf{gLastUid},\mathsf{gDocument},\mathsf{gDiffss})$. The transition rules are broken down into client side and server side rules, and are obtained from the source code~\cite{quicdoc} by manual static analysis:
\[
\setlength{\arraycolsep}{1.25pt}
\begin{array}{rl}
j_i(\epsilon,\epsilon,\epsilon)&\xrightarrow[\displaystyle{uid},\doc]{\mathsf{GET\_DOC}}j_i(uid,\doc,\epsilon)\\[16pt]
j_i(uid,\doc,\temp)&\xrightarrow[\displaystyle\diffs]{\mathsf{GET\_DIFFS},\displaystyle  uid}j_i(uid,\mathsf{applyDiffs}(\doc,\diffs),\temp)\\[16pt]
j_i(uid,\doc,\temp)&\xrightarrow{\mathsf{PUT\_DIFFS},\displaystyle uid,\mathsf{makeDiffs}(\doc,\temp)}j_i(uid,\temp,\temp)\\[16pt]
s(\luid,\doc,\diffss)&\xrightarrow[\displaystyle\luid+1,\doc]{\mathsf{GET\_DOC}}s(\luid+1,\doc,\mathsf{resetDiffs}(\diffss,\luid+1))\\[16pt]
s(\luid,\doc,\diffss)&\xrightarrow[\displaystyle\mathsf{getDiffs}(\diffss,uid)]{\mathsf{GET\_DIFFS},\displaystyle{uid}}s(\luid,\doc,\mathsf{resetDiffs}(\diffss,uid))\\[16pt]
s(\luid,\doc,\diffss)&\xrightarrow{\mathsf{PUT\_DIFFS},\displaystyle{uid},\diffs}s(\luid,\mathsf{amendDoc}(doc,uid,\diffs),\mathsf{amendDiffss}(\diffss,uid,\diffs))
\end{array}
\]
Here $\epsilon$ represents an undefined value and a side condition for all the rules is that $uid\neq\epsilon$. There is also an initial global state, which is again broken down into client side and server side parts, thus $j_i(\epsilon,\epsilon,\epsilon)$ for $i=1,2$ together with $s(0,\text{``''},[[],[]])$. Here $[...]$ represents an array.

We note in passing that breaking down the transition rules in this way provides a means of partially specifying the behaviour of each part of the application, and hence taken together they can be considered a specification of the application as a whole as evinced by its transactions over HTTP. For this reason we call these sets of rules application layer specifications.

To make use of the rules we recombine them. We combine the $\mathsf{GET\_DOC}$ rules informally first, by way of an example, generating sets of substitutions that we apply to the final states. In order to do this we note that the client side part produces a request and consumes a response whilst the server side part consumes a request and produces a response. Here the request consists of just a constant, with no other elements. The response on the other hand consists of non-constant elements. Since the client side part consumes the response, the $uid$ and $\doc$ variables of the client side part are replaced with the $\luid+1$ term and $\doc$ variable of the server side part. In order to distinguish client side elements from server side elements, subscripts are added. Once the substitutions have been made, all elements labelling the transition are discarded with the exception of the constant representing the command. The resulting global rule is the following:

\[
(j_i(\epsilon,\epsilon,\epsilon),s(\luid,\doc,\diffss))\xrightarrow{\mathsf{GET\_DOC}}(j_i((\luid+1)_s,\doc_s,\epsilon),s(\luid+1,\doc,\diffss))
\]

\begin{figure}[h]
\centering
\includegraphics[scale=1.0]{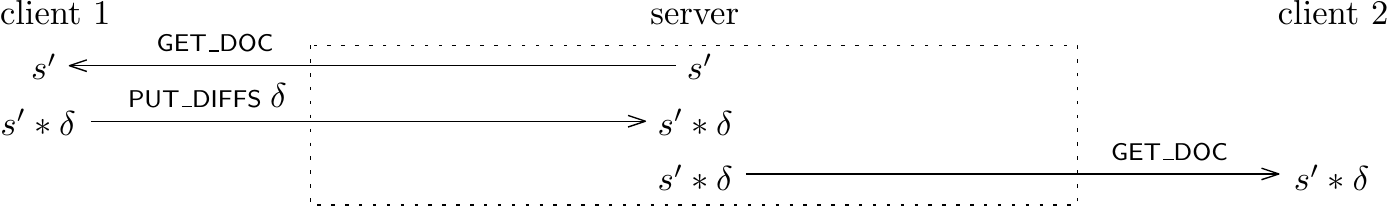}
\caption{The $\mathsf{GET\_DOC}$ and related transitions}
\label{concur_initial}
\end{figure}

\noindent We use this rule to justify an assumption in the proof of the correctness of the Concur algorithm. Specifically, lemma~6.1 in~\cite{concur} makes the assumption that the document is passed from the server to the client intact. The relevant transitions are shown in figure~\ref{concur_initial}.

\begin{lemma}
A $\mathsf{GET\_DOC}$ transition results in the $\mathsf{gWorkingDoc}$ client side variable attaining the value of the $\mathsf{gDocument}$ server side variable.
\begin{proof}
See the aforementioned global rule. The second element on the client side tuple is mapped to the $\mathsf{gWorkingDoc}$ client side variable while $doc_s$ holds the value of the $\mathsf{gDocument}$ server side variable.
\end{proof}
\end{lemma}

\noindent We next formally define a matching function for the rules in a similar vein to~\cite{DBLP:journals/tosem/FantechiGLMPT12}, collecting the elements of the requests and responses into tuples, with elements being either variables, constants or terms. A substitution is returned only if a match can be found.

\begin{definition} The partial matching function for pairing transition rules is defined by structural induction by means of the following partial functions:
\[
\setlength{\arraycolsep}{1.25pt}
\begin{array}{rcl}
\mathsf{match}((e_1,e_2...),(e'_1,e'_2...))&=&\{\mathsf{match}(e_1,e'_1)\}\cup\mathsf{match}((e_2...),(e'_2...))\\
\mathsf{match}(v,v')&=&v/v'\\
\mathsf{match}(v,c')&=&v/c'\\
\mathsf{match}(v,t')&=&v/t'\\
\mathsf{match}(c,c)&=&\empti\\
\mathsf{match}((),())&=&\{\}
\end{array}
\]
Here $e_1, e'_1, \text{etc}$ are arbitary elements, $v,v'$ variables, $c,c'$ constants and $t'$ terms. 
\end{definition}

\noindent To match the requests, we make the server side tuple the first argument of the matching function since it is this tuple that contains the variables that must be replaced in the substitutions that follow. To match the responses, we make the client side tuple the first argument. By way of an example we show the matching of the requests and responses of the $\mathsf{GET\_DIFFS}$ rules:
\[
\setlength{\arraycolsep}{1.25pt}
\begin{array}{rl}
\setlength{\arraycolsep}{1.25pt}
\begin{array}{cl}
&\mathsf{match}((\mathsf{GET\_DIFFS},uid_s),(\mathsf{GET\_DIFFS},uid_i))\\
=&\{\empt\}\cup\mathsf{match}((uid_s),(uid_i))\\
=&\{\empt\}\cup\{\mathsf{match}(uid_s,uid_i)\}\\
=&\{\empt\}\cup\{uid_s/uid_i\}\cup\mathsf{match}((),())\\
=&\{\empt,uid_s/uid_i\}
\end{array}
&
\setlength{\arraycolsep}{1.25pt}
\begin{array}{cl}
&\mathsf{match}((\diffs_i),(\mathsf{getDiffs}(\diffss_s,uid_s)_s))\\
=&\{\mathsf{match}(\diffs_i,\mathsf{getDiffs}(\diffss_s,uid_s)_s)\}\cup\mathsf{match}((),())\\
=&\{\diffs_i/\mathsf{getDiffs}(\diffss_s,uid_s)_s\}\\
\\
\\
\end{array}
\end{array}
\]
Here the rules are paired as expected by matching the constants, and we apply the substitutions to obtain the final states of the global rule:
\[
\setlength{\arraycolsep}{1.25pt}
\begin{array}{rcl}
&&j_i(uid,\mathsf{applyDiffs}(\doc,\diffs),\temp).\{\diffs_i/\mathsf{getDiffs}(\diffss_s,uid_s)_s\}.\{\empt,uid_s/uid_i\}\\
&=&j_i(uid,\mathsf{applyDiffs}(\doc,\mathsf{getDiffs}(\diffss_s,uid_s)_s),\temp).\{\empt,uid_s/uid_i\}\\
&=&j_i(uid,\mathsf{applyDiffs}(\doc,\mathsf{getDiffs}(\diffss_s,uid_i)_s),\temp)
\end{array}
\]
\[
\setlength{\arraycolsep}{1.25pt}
\begin{array}{rcl}
&&s(\luid,\doc,\mathsf{resetDiffs}(\diffss,uid)).\{\empt,uid_s/uid_i\}.\{\diffs_i/\mathsf{getDiffs}(\diffss_s,uid_s)_s\}\\
&=&s(\luid,\doc,\mathsf{resetDiffs}(\diffss,uid_i)).\{\diffs_i/\mathsf{getDiffs}(\diffss_s,uid_s)_s\}\\
&=&s(\luid,\doc,\mathsf{resetDiffs}(\diffss,uid_i))
\end{array}
\]

\begin{figure}[h]
\centering
\includegraphics[scale=1.0]{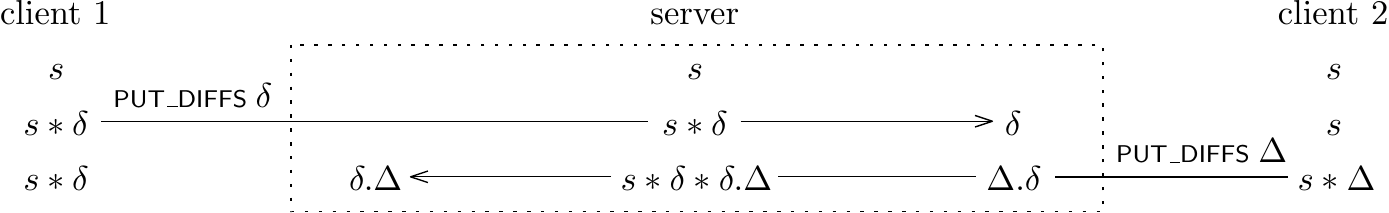}
\caption{The $\mathsf{PUT\_DIFFS}$ transitions}
\label{concur_base}
\end{figure}

\noindent Note that both substitutions must be applied to the final states, since terms labelling the transitions may themselves require subsitutions. For example, the $\{\empt,uid_s/uid_i\}$ substitution can be applied to the $\mathsf{getDiffs}(\diffss_s,uid_s)_s$ term to yield $\mathsf{getDiffs}(\diffss_s,uid_i)_s$, which can then be substituted into the final state of the client side rule, yielding the same result. We choose to apply the substitutions in reverse order so that they can both be applied directly to the final state, however.

\begin{lemma}
A $\mathsf{GET\_DIFFS}$ transition results in the client in question applying its pending diffs to its copy of the document, held in the $\mathsf{gWorkingDoc}$ client side variable. The client in question's pending diffs are then reset.
\begin{proof}
See the aforementioned final states above.
\end{proof}
\end{lemma}

\noindent To conclude we consider the global rule for the $\mathsf{PUT\_DIFFS}$ command. In this case we apply the requisite substitutions but keep the client side and server side parts separate due to space considerations. We treat the client side part first:
\[
j_i(uid,\doc,\temp)\xrightarrow{\mathsf{PUT\_DIFFS}}j_i(uid,\temp,\temp)
\]
The implementation maps the $\mathsf{gTempDoc}$ client side variable directly to the textarea. When its value changes, diffs are generated from the difference between this variable and the $\mathsf{gWorkingDoc}$ variable. When these diffs are put on the server, the $\mathsf{gWorkingDoc}$ variable is updated with the value of the $\mathsf{gTempDoc}$ variable. The above rule simply states this fact. Moving on, we give the server side part of the global rule:
\[
s(\luid,\doc,\diffss)\xrightarrow{\mathsf{PUT\_DIFFS}}s(\luid,\mathsf{amendDoc}(doc,uid_i,\diffs_i),\mathsf{amendDiffss}(\diffss,uid_i,\diffs_i))
\]
Here we have abbreviated $\mathsf{makeDiffs}(\doc_i,\temp_i)_i$ as $\diffs_i$. We use this rule to justify another assumption in the proof of the Concur algorithm. Specifically, lemma~6.2 in~\cite{concur} makes the assumption that if clients put diffs on the server, these are amended appropriately and applied to its copy of the document. The relevant transitions are shown in figure~\ref{concur_base}.

\begin{lemma}
A $\mathsf{PUT\_DIFFS}$ transition results in the $\mathsf{gDocument}$ server side variable being updated with the return value of the $\mathsf{appendDoc}(...)$ server side method, whose arguments are the identifier and diffs passed by the client in question.
\begin{proof}
See the aforementioned global rule, where the second element of the server side tuple is mapped to the $\mathsf{gDocument}$ server side variable while $uid_i$ and $\diffs_i$ hold the values of the $\mathsf{gUid}$ client side variable and return value of $\mathsf{makeDiffs}(...)$ client side method, respectively.
\end{proof}
\end{lemma}

\section{Related work}

The approach taken in~\cite{DBLP:conf/csfw/AkhaweBLMS10} defines a formal model of web security and identifies, amongst others, vulnerabilities in WebAuth, a web-based authentication system based on Kerebos. The concept of non-linear time is adopted, with the authors not being ``...concerned with the actual temporal order between unrelated actions'', an approach strongly espoused here. 

In~\cite{DBLP:conf/kbse/LicataK04}, significant emphasis is placed on user interactions. The opinion that ``...any verfication tool for the web that does not account for user operations is not only incomplete, but potentially misleading'' could be construed as being somewhat the opposite to our own. In~\cite{DBLP:journals/jcss/DeutschSV07}, emphasis is placed on the structure of the web site in terms of its interelated links, with transitions being seen as taking place between web pages. In effect a temporal order is imposed, with users choosing only from a set of actions presribed by each web page, again in marked contrast to the approach taken in~\cite{DBLP:conf/csfw/AkhaweBLMS10} and our own.

By eschewing a model of user interactions, our approach owes more to the analysis of web services~\cite{DBLP:conf/isola/RavnSV10} than applications. And although our approach is not a model checking one, it has close parallels with the exhaustive treatments that typify model checking. Lastly, we note that our approach bears some resemblance to the inductive analysis of network protocols~\cite{DBLP:journals/tissec/Paulson99}, which effectively build an operational semantics of protocols given a set of rules and then prove that certain safety properties are preserved.
\vspace{-0.5em}

\section{Conclusions and future work}

In this paper we have developed a formal model of HTTP transactions and Internet application state. We have defined two formal, example applications based on this model and proved properties of these applications. We then extended our approach to the treatment of a case study, namely a rich, distributed Internet application, and proved properties of this application, too. Because our model is founded only on HTTP transactions and Internet application state, and not on any particular framework or programming paradigm, we claim that it can be used to prove properties of any Internet program or protocol. 

The case study chosen was relatively simple, with both server side and client side parts written in JavaScript. Our long term goal is more ambitious, however. In future work we plan to use application layer specifications to formalise properties of larger HTTP-based programs and protocols with, specifically, the OAuth 2.0 protocol~\cite{oauth} in mind.

The choice of the logic UCTL* was motivated by the need for formulae with nested next time operators but, rather than defining the $\phi T_\chi\phi'$ operator as shorthand for $\phi\wedge X_\chi\phi'$, we could define its semantics directly and otherwise do away with the need for a fully branching logic. Furthermore, as demonstrated in the case study, it is possible to express properties of these systems without using a logic. For this reason, developing further variants of UTCL* or any other logic is currently not the focus of future work.

Finally, we make some comments on the use of terms in the transition rules in the case study. The treatment was somewhat informal, however a formal treatment can be found for example in~\cite{Rutten94initialalgebra}, which includes transition systems generated by rules, sets of which are gratifyingly called transition system specifcations. Moreover, semantics that encompases both the denotational meaning of terms and their operational meaning over transition systems are developed therein. We conclude therefore by looking forward to a formal treatment of the ideas presented here based on these approaches.

\subsection{Acknowledgements}

The author thanks the anonymous reviewers for their many helpful comments and corrections.

\nocite{*}
\bibliographystyle{eptcs}
\bibliography{references}

\end{document}